# The Rise of Null Hypothesis Significance Testing (NHST): Institutional Massification and the Emergence of a Procedural Epistemology

Carol Ting[*]

Department of Communication, University of Macau

Note. This paper examines the rise of Null Hypothesis Significance Testing (NHST) from a historical and sociological perspective. While engaging with statistical and inferential debates, the analysis focuses on the institutional conditions under which particular inferential technologies stabilize and spread.

## *Abstract*

It has long been a puzzle why, despite sustained reform efforts, many applied scientific fields remain dominated by Null Hypothesis Significance Testing (NHST), a framework that dichotomizes study results and privileges “statistically significant” findings. This paper examines that puzzle by situating the development and rise of NHST within its historical and institutional context. Taking Actor-Network Theory as a point of entry, the analysis identifies the conditions under which particular inferential technologies stabilize and endure. The analysis shows that, although NHST does not resolve the technical problem of statistical inference, it came to dominate as a social technology that addressed the most pressing institutional challenge of the postwar period: the mass expansion of scientific networks. Under conditions of rapid institutional growth, NHST’s technical slippages—purging research context and replacing epistemic judgment with mechanical procedures—became functional features rather than flaws. These features enabled procedural self-sufficiency across settings marked by heterogeneous goals and uneven expertise, thereby sealing NHST’s position as the obligatory passage point in many postwar scientific fields.

---

[*] tingyf@gmail.com

## Introduction

"Research has shown…" is a phrase that carries almost mystical authority in modern society. When we question those research findings, replies often conclude with the inequality "$p \leq 0.05$" and the term "statistically significant". These case-closing terms are the output of Null Hypothesis Significance Testing (NHST hereafter), the dominant statistical testing method taught in most entry-level applied statistics courses in the social and biomedical sciences.

Journal editors and reviewers often rely on whether the $p$-value falls below 0.05 to determine whether submitted studies offer new insights. Researchers, media reporters, decisionmakers at funding agencies, and managers at academic institutions, in turn, rely on a researcher's journal publication record to judge the quality and impact of her work. In this sense, the $p$-value has come to function as the currency in the empire of NHST.

Despite its ubiquity and authority in adjudicating scientific claims, NHST has been the target of criticism and statistical reform for more than sixty years (Amrhein, et al., 2019; Bakan, 1966; Cohen, 1994; Gelman, 2016; Greenland, et al., 2016; Sterling, 1959; Wasserstein & Lazar, 2016; Wasserstein, et al., 2019). Oakes (1986) was among the first to document common misconceptions about the $p$-value, based on a sample of British academic psychologists. Subsequent studies have reported similar patterns across different populations. Haller and Krauss (2002), for example, found comparable misunderstandings among German professors and students in psychology. More recently, related findings have been reported in Spain, Italy, Netherlands, and Chile (Badenes-Ribera, Frias-Navarro, Monterde-i-Bort, & Pascual-Soler, 2015; Badenes-Ribera, et al., 2016; Hoekstra, et al., 2014). Examining journals in medicine, cognition, psychology, business, and economics, McShane and Gal (2016) suggest that authors—as well as editorial board members –often struggle to interpret $p$-values appropriately. McShane and Gal (2017) further note that such difficulties are not confined to non-statisticians.

Given modern society's reliance on scientific evidence, how can we account for the prevalence and persistence of misconceptions about NHST despite decades of calls for reform and a large body of literature documenting its flaws? Drawing on work in the history of statistics and writings by statisticians, this paper examines the emergence and subsequent dominance of NHST. This perspective makes it possible to consider how its rise may be connected to what Gigerenzer et al. (1989, p. 106) describes as a chimerical ancestry—the offspring of a "forced marriage" between two theoretically incompatible frameworks. The two sides fought long and hard, but as their ideas disseminated through the rapid postwar expansion of scientific networks, neither remained intact.

I argue that massive post war institutional expansion (massification) of American higher education—marked by unprecedented growth in science-related fields and sustained government investment in science education—created conditions under which methods capable of scaling on demand were particularly attractive. From this vantage point, NHST's success can be understood in relation to its capacity to purge context and operate as a procedurally self-sufficient method, an argument developed through the historical and sociological analysis that follows. In this sense, NHST can be viewed as a social technology of institutional massification, whose technical slippages facilitated its circulation and endurance across diverse settings.

This study takes Actor-Network Theory (ANT, Latour, 1987; Callon, 1986) as a point of entry. Concepts of black box and the translation model of fact-building are useful for analyzing the emergence and spread of NHST. At the same time, by anchoring the NHST network in the postwar period of rapid expansion of science, this study foregrounds the role of institutional incentives and demographic change, situating these processes within a broader structural and historical context in which particular methodological forms were selected and stabilized.

The concept of "a technology of institutional massification" may invite comparison with Theodore Porter's notion of a "technology of trust" (1995). While Porter emphasizes how quantification enables disciplines and professions to defend their autonomy against outside scrutiny, the focus here is on why particular forms of quantified procedure—such as NHST—came to dominate across diverse disciplines and continues to attract new followers. This paper develops the concept of procedural self-efficiency to capture one mechanism through which NHST is able to travel without the need for repeated epistemic renegotiation, and to become intertwined with postwar processes of institutional massification.

The analysis also draws on Gerd Gigerenzer's work on the history of probability and statistics (Gigerenzer, et al., 1989), critiques of NHST (2004; 2015), and discussion of its role in the replication crisis (2018). Gigerenzer has characterized NHST as a surrogate for scientific research (2015) and a social ritual that shelters researchers from the responsibility of making substantive judgments (2018). By foregrounding the social and historical conditions under which NHST emerged, this paper builds on these insights while shifting attention from individual cognition toward institutional context. From this perspective, correcting cognitive errors and improving statistical education remain part of the story, but attention to institutional context become indispensable. The sections that follow trace this process from the early development of statistical testing to its consolidation under postwar conditions of expansion.

The paper proceeds as follows. The next sections reconstruct the technical and conceptual foundations of hypothesis testing, tracing the distinct inferential logics developed by Fisher and by Neyman and Pearson and the institutional conditions under which these frameworks were translated, simplified, and ultimately fused. Readers who are already familiar with the technical literature on significance testing may safely skim or skip these sections and move directly to the historical analysis of postwar expansion and institutional massification, where the central sociological argument is developed. The latter sections examine how NHST stabilized as a procedurally self-sufficient method under conditions of rapid growth in scientific research, and conclude by reflecting on the implications of this history for contemporary debates over statistical reform.

## A Brief History of Null Hypothesis Significance Testing

Historical accounts trace the early history of probability and statistics to at least the seventeenth century, when mathematical-inclined researchers took interest in games of chance and sought rational strategies for dealing with patterned uncertainty (Stigler, 1986). During the Enlightenment, mathematicians recognized the broader potential of this line of inquiry, and the study of games of chance soon found applications in both the natural sciences (such as astronomy and physics) and the social sciences (including actuarial science and population statistics).

While the large-scale adoption of statistics took place mainly in the twentieth century, tools and concepts resembling modern statistical methods were already in use in the nineteenth century (Shafer, 2020; Stigler, 1986). The early twentieth century saw Karl Pearson's breakthrough work on the goodness-of-fit test (K. Pearson, 1900) and William Gosset's seminal development of the small sample distribution later used in *t*-test (Student,[1] 1908). However, prior to the work of Ronald A. Fisher (1890-1962), there was no unified logical framework for reasoning from sample to population: the concepts of population and sample were not yet clearly distinguished, and statistical tools were often used in an *ad hoc* and domain-specific manner (Rao, 1992; Box, 1978). Contemporary commentators, including Fisher himself, noted that this lack of conceptual clarity hindered the development of statistical inference (Fisher, 1922).

At the same time, given the inherent difficulty of inferential reasoning, forms of misuse and misconceptions were already common. As early as 1843, for example, the economist-mathematician Cournot documented practices that bear a close resemblance to what would now be described as p-hacking[2] (Shafer, 2020).

### ***R. A. Fisher's Inductive Inference: Reasoning from Sample to Population***

Ronald A. Fisher opened a new page in statistics with his "On the Mathematical Foundations of Theoretical Statistics" (1922). In this seminal work, he clearly defined and distinguished the concepts of population and sample, and from there he laid out theoretical considerations for using a sample to model and learn about the world. Here, the *sample* refers to the observed data, while the *population* denotes the entire class of objects from which the sample is randomly drawn.

Fisher points out that statistical work in general involves three types of problems (Fisher, 1922, p. 313):

(1) Specification. In inferring about a population from samples, statisticians first use their judgment to narrow down possibilities by specifying a reasonably accurate population distribution that can be summarized by a small number of parameters, such as a normal distribution (which is uniquely identified by two parameters: center and dispersion). The choice of a population distribution is an empirical matter and requires expert judgment. At Fisher's time, computers and simulations were unavailable, so this choice was necessarily confined to "those forms which we know how to handle, or for which any tables which may be necessary have been constructed" (p. 314). Much of the groundwork for later ideas of distributional specification had been laid before the twentieth century. For example, Gauss and Laplace showed that repeated measurement errors follow a normal distribution, while Galton later found similar bell-shaped patterns in human traits, especially height. Other well-known distributions, such as the Poisson and binomial, were also in use in particular domains (Stigler, 1986). Once a population distribution is specified, the relationship between population and samples can be studied on that premise, which involves the other two interwoven tasks.

---

[1] Gosset worked as a chemist at the Guiness Company in Dublin when he published the paper. He used the pseudo name "Student" to avoid violating Guiness' policy.

[2] This is a practice where researchers take advantage of the decisions made during the research process to produce *p*-values under the 0.05 threshold, which is implicitly required for journal publication (Simmons, et al., 2013).

(2) Estimation. Constructing a method that uses sample data to produce estimates of a given population parameter.

(3) Distribution. Deriving the distributional pattern of sample estimates.

Specification is discussed in greater detail here because it makes explicit the inferential setup that estimation and the derivation of sampling distributions presuppose, which are taken up in the following section.

Fisher's major contribution lies in his work on estimation methods and the distributions of sample estimates. He defines a *parameter* as a numerical summary of a specific aspect of a population—what we want to know but cannot directly observe. Examples include the average blood pressure of 50-year-old females, the proportion of voters voting for a candidate, or the effect of weight-loss drug for a specific demographic group. While a parameter is a population concept, its sample counterpart is called a *statistic*; in other words, a statistic is a sample estimate of a population parameter. To emphasize this distinction, this paper often uses the term sample estimate in place of statistic.

The relationship between sample statistics and a population parameter is often illustrated with the analogy of shadows and the object that casts them. Just as shadows hint at the object, sample estimates hint at the population parameter. If we record the shadow cast on a wall by an outdoor object over a few sunny days, the envelope of those shadows provides a record of their regular behavior. If we then observe a shadow extending beyond that envelope, we may infer that the shadow is cast by a different object, or that the original object has been moved. In this example, inference seems simple and intuitive because it draws on decades of embodied experience with light, objects and shadows. This contextualized form of inference is deeply internalized and has become largely invisible to us, to the point that it is taken for granted. Babies—who have not yet acquired this background knowledge—are often baffled by shadows. Connecting a sample statistic to a population parameter likewise requires contextual understanding. Unlike shadows, however, statistical inference is technically complex and depends on modeling assumptions that are not always transparent to users, making it particularly vulnerable to misunderstanding.

### ***Sampling Distribution: The Bridge Between Sample Estimates and Population Parameters***

To better understand the problem associated with NHST and what contributes to its appeal, we need to delve into the concept of the *sampling distribution*, which links sample estimates to a given population parameter. Due to random variation, some samples contain more extreme cases than others, so each sample produces a different estimate. A sampling distribution can be understood as an approximate description of how study results may vary across repeated samples. A tall and narrow bell-shaped distribution indicates that sample estimates are more likely to fall close to the center of the distribution, while values in the tail areas are far less likely to be observed (See Figure 1 for an illustration). Everything else being equal, a narrower sampling distribution is generally preferred because it implies that sample outcomes are less variable and more stable. Because the sampling distribution describes the behavior of sample estimates, work on estimation method and the derivation of sampling distributions is typically carried out together.

More specifically, a sampling distribution describes the range of possible values of sample estimates and their corresponding frequency or probability under a specified statistical model of the data-generation process. The data-generation process itself includes all contextual factors affecting possible sample outcomes—that is, study design, treatments administered, sampling method, sample size, measurement method, and so on. To determine how a population parameter can be estimated and how sample estimates will behave, statisticians must make judgments about how the data are generated. These judgments identify which aspects of the data-generation process are most relevant and translate them into explicit modeling assumptions. Under these assumptions, statisticians formulate a sample estimator—defined as a mathematical function of the sample data—from which the sampling distribution can then be derived.

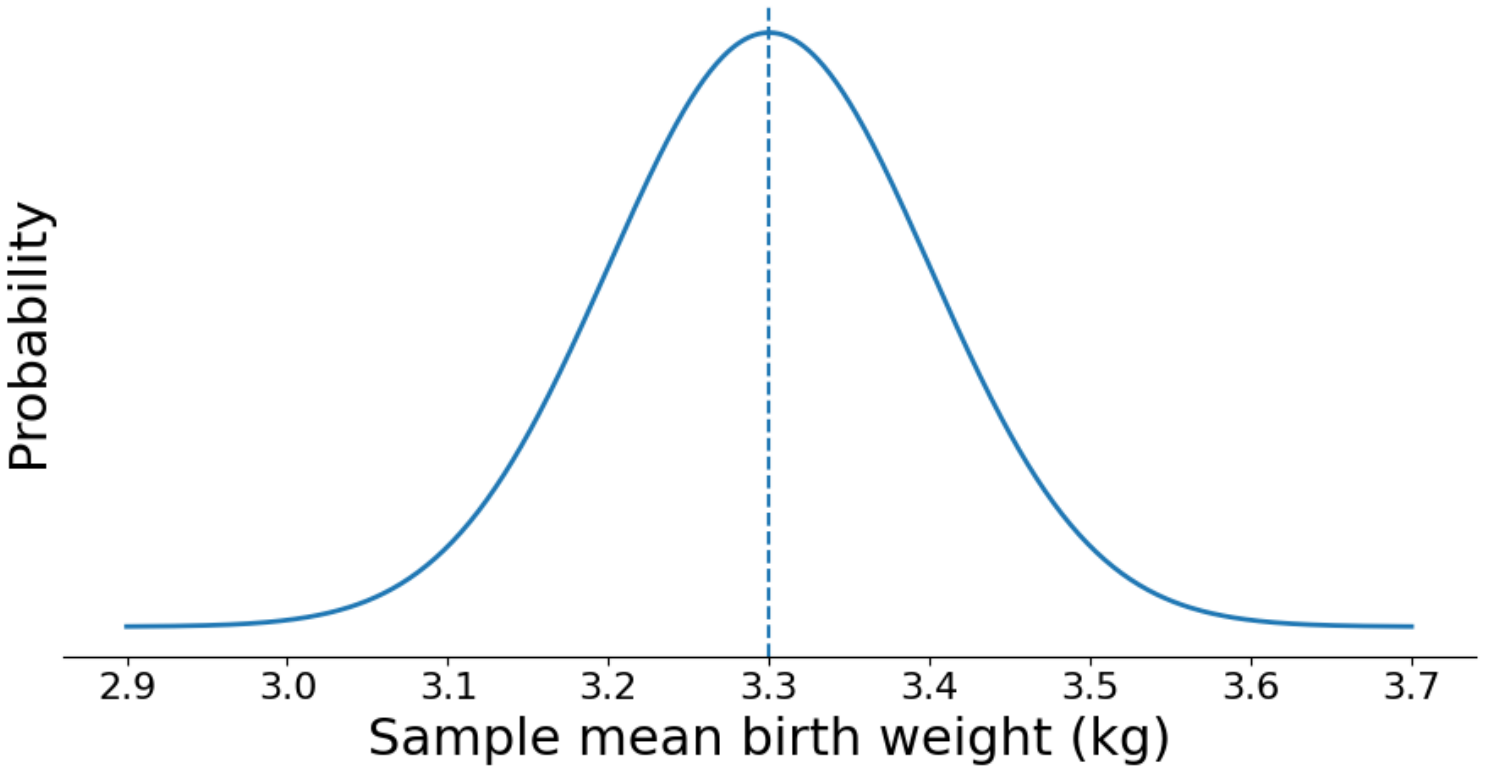

Figure 1 A bell-shaped sampling distribution of mean birth weight

Modeling assumptions are therefore fundamental to estimators and sampling distributions. If we think of the estimator and its sampling distribution as the bridge that connects sample to population, the modeling assumptions are the piers holding up the bridge. Violations of the assumptions can lead to systematic bias in sample estimates, and in serious cases render results unreliable or even useless. Some examples of commonly invoked modeling assumptions include random sampling in observational studies, random assignment to experimental conditions, independence and identical distribution of random variables, and data reporting practices that are not contingent on *p*-values. The list extends further to assumptions specific to particular tests. In other words, every statistical test rests on its own set of modeling assumptions, and the validity of its results depends on the extent to which those assumptions are met.

Due to technical complexity, it is not easy to show or explain how violations of different modeling assumptions affect the validity of statistical outputs. The effect of sample size on the sampling distribution is one of the few factors that can be readily calculated and demonstrated, and textbooks do tend to emphasize its impact. In contrast, the effects of many other assumptions—which may be equally or even more important—are routinely omitted, to the point that most students are unaware of them or come to regard them as peripheral. As we will see later, this pattern of omission plays an important role in understanding the rise and dominance of NHST.

***Fisher's Significance Testing: What to Make of the Study Results?***

In addition to estimation, Fisher also developed testing procedures that he called "significance tests", which are often conflated with NHST. Although NHST has a Fisherian lineage, the two rest on markedly different logical foundations.

Between 1919 and 1933, Fisher worked at the Rothamsted Agricultural Experimental Station. The ample experiment data and practical problems encountered at Rothamsted served as a catalyst for some of his most innovative ideas in statistics and experiment design. Researchers there faced a recurring question: what should one make of the result of a novel study?" For genuinely novel research, there is often little prior information to rely on, and learning must proceed through the elimination of possibilities in a manner consistent with a logic of falsification. A researcher studying the effect of a fertilizer, for example, might begin with the test hypothesis that "this fertilizer does not improve harvest"—that is, that the difference between outputs from fertilized and unfertilized plots is zero. Fisher called this test hypothesis the null hypothesis, $H_0$, and treated it as a hypothesis about a population parameter. Importantly, the test hypothesis need not be zero. For instance, in drug testing researchers may be more concerned that the side effects do not exceed a given tolerable level. In the following discussion, $H_0$ refers to this more general sense of a test hypothesis.

When all modeling assumptions hold, the $H_0$ sampling distribution is centered on the hypothesized effect size (Figure 2), and its dispersion—typically operationalized as the standard error of the estimate—can be estimated from sample data. To test the hypothesis, the observed sample estimate is compared to this $H_0$ distribution. The intuition is straightforward: a sample estimate far from the center of the $H_0$ distribution corresponds to a low-probability event, and the degree of improbability serves as evidence against $H_0$. Crucially, the evidential weight does not lie in the observed effect size, but in how extreme that estimate is relative to the sampling variability expected under $H_0$. In Fisher's testing framework, the tail areas of the $H_0$ distribution sum up the probability of observing outcomes at least as extreme as the observed sample estimate, and this quantity is the *p*-value (Figure 2).

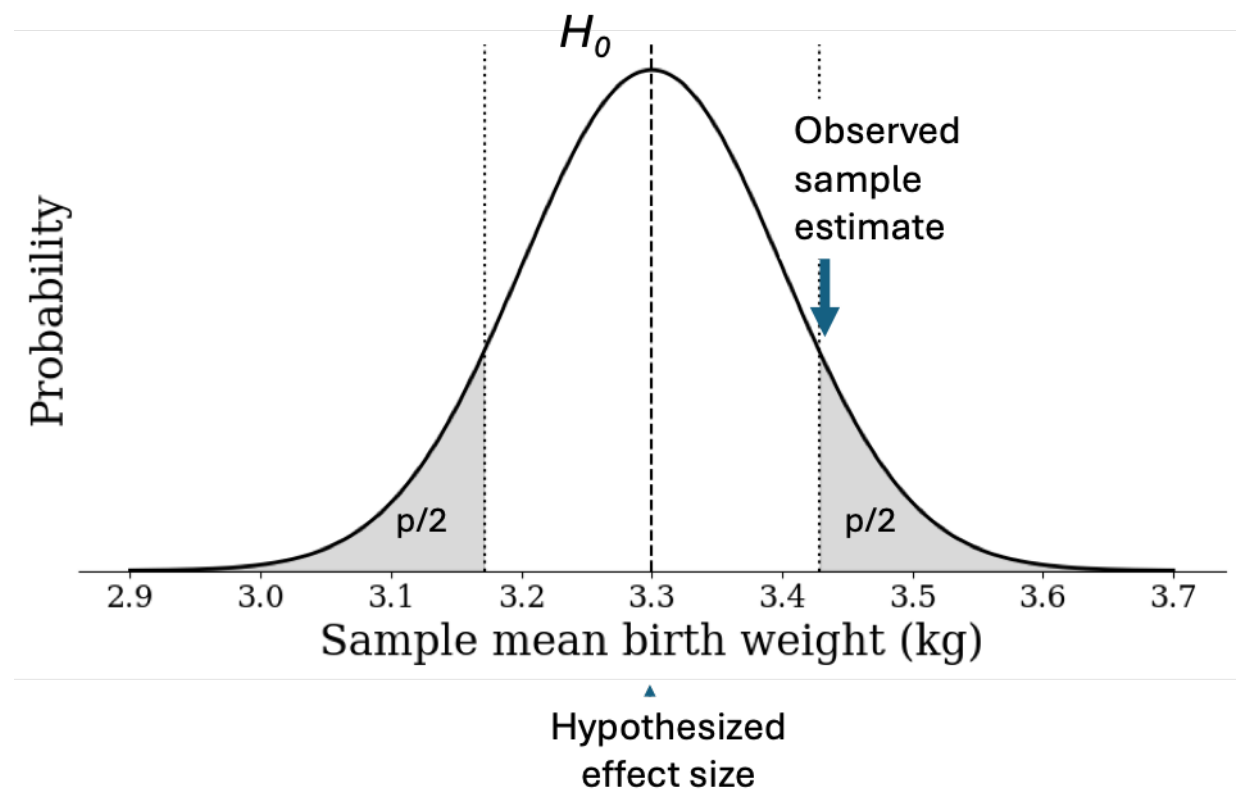

Figure 2 $H_0$ and the *p*-value in the Fisherian framework

It is important to emphasize that, in Fisher's framework, the *p*-value carries evidential value. This is because the smaller the *p*-value, the larger the gap between the hypothesis and the data, and the more surprised the observed results would be if $H_0$ were true. This naturally raises the question of how small the *p*-value should be to warrant rejecting $H_0$ as implausible. In his earlier writings, Fisher used a 0.05 threshold as a rule-of-thumb. Results with *p*-values below this threshold came to be described as "statistically significant", while those above it were labeled statistically non-significant. In his latter works, however, Fisher repeatedly emphasized that no fixed threshold could substitute for scientific judgment, and that appropriate significance levels depends on the context of inquiry: "no scientific worker has a fixed level of significance at which from year to year, and in all circumstances, he rejects hypotheses; he rather gives his mind to each particular case in the light of his evidence and his ideas" (Fisher, 1956, p.42).

As explained earlier, the sampling distribution—and therefore the *p*-value—depends on sample size. Everything else being equal including the size of the observed departure from $H_0$, a larger sample size produces a narrower sampling distribution and thus a smaller *p*-value. This is because increasing sample size reduces sampling variability, which makes the observed estimate appear farther from the hypothesized value relative to that variability (Greenland, 2019). In other words, the *p*-value reflects both the estimated effect size and the estimated variability of such estimates. This is entirely reasonable, but it does complicate interpretation: when sample size is very large, even substantively trivial effect size can yield statistically significant results. This tension has contributed to persistent confusion, as readers often equate "statistical significance" with practical importance and thereby overstate the impact of findings.

Fisher calls his approach "inductive inference" (1935b), in the sense that inferences made about a population are based on induction from sample data. This framework is compatible with the Popperian logic of falsification, according to which theories can be falsified but never definitively confirmed. Within this framework, rejection based on a small *p*-value means that the gap between $H_0$ and the data is large enough to cast $H_0$ in doubt—provided that all modeling assumptions are met. This signals that the researcher may be onto something, but that further investigation is warranted (Fisher, 1935a, p. 16). In contrast, a *p*-value above the conventional threshold does not say anything about the true effect size; it merely indicates that the data are reasonably compatible with $H_0$ and the modeling assumptions—a very low bar. Such results can therefore only be regarded as inconclusive. This asymmetry follows directly from Fisher's setup: the tests and its threshold are designed to probe the plausibility of $H_0$, not to evaluate alternative effect sizes. For this reason, *p*-values above the threshold do not provide information about the magnitude of the true effect. Acceptance of $H_0$ is not possible within this inferential framework.

***Neyman-Pearson Testing Framework: How to Make Optimal Long-Run Decision Rules?***

A pair of theoretical statisticians, Jerzy Neyman and Egon Pearson (son of Karl Pearson), approached hypothesis testing in a very different way. They found single-hypothesis tests unsatisfactory, arguing that such tests inevitably rely on researchers' judgment without providing a principled basis for decision-making. As they put it:

> [a]ll that is possible for [an investigator] is to balance the results of a mathematical summary, formed upon certain assumptions, against other less precise impressions based upon a priori or a' posteriori considerations. The tests them-selves give no final verdict, but as tools help the worker who is using them to form his final decision; one man may prefer to use one method, a second another, and yet *in the long run* there may be little to choose between the value of their conclusions. (Neyman & Pearson, 1928, p.176) [Emphasis added]

Neyman and Pearson criticized single-hypotheses tests as imprecise and arbitrary because they depend on cutoff values chosen by researchers on a case-by-case basis. As discussed earlier, such tests face a structural problem: with sufficiently large samples, even very small departures from $H_0$ can produce *p*-values below a chosen threshold, making rejection possible for virtually any effect size.

For theorists like Neyman and Pearson, this is not rigorous enough. They sought automatic decisions based on rules that could be mathematically justified. Responding to what they saw as the arbitrariness of single-hypothesis testing, Neyman and Pearson proposed a different testing framework, in which they

> …look at the purpose of tests from another view-point. *Without hoping to know whether each separate hypothesis is true or false*, we may search for rules to govern our behaviour with regard to them, in following which we insure that, *in the long run* of experience, *we shall not be too often wrong… Such a rule tells us nothing as to whether in a particular case H is true when* $x \leq x_0$ *or false when* $x > x_0$. But it may often be proved that if we behave according to such a rule, then *in the long run* we shall reject H when it is true not more, say, than once in a hundred times, and in addition we may have evidence that we shall reject H sufficiently often when it is false. (Neyman & Pearson, 1933, p.291) [Emphasis added]

These statements make clear that Neyman and Pearson were tackling a fundamentally different problem. Rather than drawing inference from individual samples, they were willing to trade case-by-case judgment for long-run control of error rates. In other words, their target was to formulate tests and decision rules whose performance could be guaranteed over repeated use, a perspective that has come to be known as the decision-theoretic framework. Despite their explicit distinction from Fisher's approach, confusion between the two testing frameworks quickly emerged and became stabilized in subsequent practice.

To arrive at a mathematically rigorous optimal decision rule, Neyman and Pearson introduced an alternative hypothesis symmetric to the original test hypothesis. For expositional clarity, let's call the original test hypothesis $H_0$ and the alternative $H_1$. Here "symmetric" means that $H_1$ should be specified as precisely as $H_0$, with a particular center, such as "the effect of the drug is to reduce blood pressure by 10 mmHg". Hypothesis testing is thus reframed as a choice between two clearly specified hypotheses, with the sample used to decide between them. Figure 3 illustrates this choice.

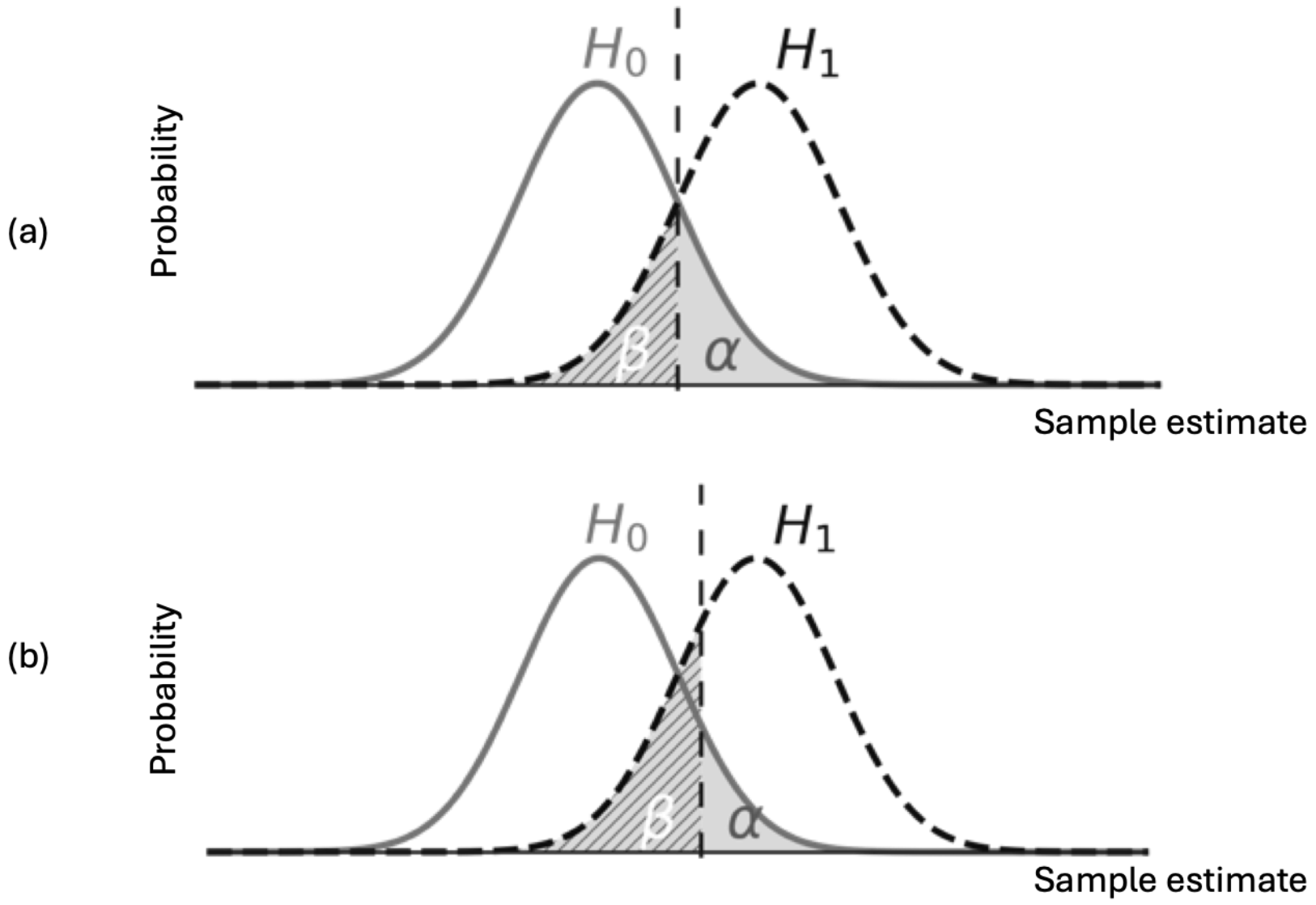

Figure 3 Illustration of Neyman and Pearson's decision theoretical framework

As long as the two sampling distributions overlap, errors are unavoidable. No matter where the cutoff is placed, some samples generated under $H_0$ will fall on the $H_1$ side of the cutoff, and some samples generated under $H_1$ will fall on the $H_0$ side. (Here $H_1$ is assumed, without loss of generalizability, to lie to the right of $H_0$.) Comparison of panels (a) and (b) in Figure 3 shows that the choice of cutoff determines the probability of wrong decisions. When a larger cutoff is chosen (Panel b), fewer outlying sample estimates from the $H_0$ distribution—those falling to the right of the cutoff—will lead to false rejection of $H_0$. Such errors are called Type I errors, and their probability is marked by the area $\alpha$. Reducing the Type I error rate necessarily increases the probability of Type II errors (marked by the area $\beta$), in which outlying samples from the $H_1$ distribution lead to false rejection of $H_1$.

The tradeoff between Type I and Type II error rates turns hypothesis testing into an elegant mathematical optimization problem—one that proved especially attractive within theoretical statistics and well suited for formal analysis and training. Neyman and Pearson showed that, for many cases, there exists a most "efficient" decision rule that minimizes the Type II error rate for a given Type I error rate.

This mathematically elegant construction, however, comes with demanding information requirements that are rarely met on the genuine research frontier. If a study is genuinely novel, the researcher typically has little prior information on which to specify $H_1$, let alone to determine an appropriate value for $\alpha$. As a result, practical applications of the Neyman-Pearson framework have largely been confined to settings in which sufficient prior information has accumulated to support the specification of optimal decision rules, such as in industrial quality control and risk management (Gigerenzer, et. al, 1989, pp. 101-102).

An important and distinct aspect of the Neyman-Pearson approach is the role of the *p*-value, which takes on a fundamentally different meaning than in the Fisherian framework. Although it is also a continuous value between 0 and 1, in the Neyman-Pearson framework the *p*-value functions as a random outcome used to choose between $H_0$ and $H_1$. It plays a role analogous to that of a coinflip in a random decision procedure. Just as, before flipping a coin, we must specify rules such as Heads-I-pay, Tails-you-pay, before drawing a sample to decide between $H_0$ and $H_1$ we must set a rule such as: if $p$ exceeds $\alpha$, reject $H_0$; otherwise, accept $H_0$. In this framework, neither the outcome of the coinflip outcome nor the *p*-value carries evidential value. They cannot tell us anything about the probability that the decision is correct in a particular case. The decision is either right or wrong, and the method itself is silent on this matter.

To sum up, the Neyman-Pearson framework achieves two things. First, by recontextualizing hypothesis testing as an optimization problem of making automatic decisions, it allows for control of long-run error rates. Second, by introducing a clearly specified alternative hypothesis, Neyman and Pearson make it possible to "accept" $H_0$—without contradicting the Popperian notion that theories can only be falsified, not proven. Importantly, these achievements come with significant costs. The demanding information requirements limit the framework's practical applicability, especially at the genuine research frontier. Moreover, outcomes of individual studies have no evidential value in the Neyman-Pearson context. As the quotations above make clear, any individual decision made under this framework may be wrong, and there is no way to know the probability of error in a particular case. The decision rule guarantees only that, in the long-run, error rates will be controlled at pre-specified levels.

## The Pre-war Years: Institution Building in Elite Circles

In terms of institution building, the prewar years were a slow period for the discipline of statistics. Up until the mid 1930s, statistics was highly mathematical and largely confined to elite circles. Despite the heated and polemical debate between Fisher and Neyman-Pearson, the substantive issues were too abstract and technically demanding to travel widely beyond those circles or to be incorporated into routine scientific training. As a result, statistical theory developed largely within small networks of specialists, with limited institutional capacity for large-scale teaching or diffusion across disciplines.

In 1925, Fisher published *Statistical Methods for Research Workers* ([SMRW], Fisher, 1925) as a practical manual for researchers. In it, Fisher guides users through common inferential tasks with worked examples showing how his method could be applied to data. To facilitate significance testing, the book also provides tables of critical values for test statistics, allowing researchers to determine whether their results met chosen cutoff thresholds for significance. Although it was immediately recognized for its innovations and contribution (Hotelling, 1927), the book was also criticized for leaving out formal proofs and for a writing style that is "at once polished and award" (Savage, 1976, p. 448).

It took George W. Snedecor of Iowa State College of Agriculture (now Iowa State University, ISU) to "act as midwife in delivering the new statistics in the United States" (Box, 1978, p. 313). A mathematician working on statistical applications in agriculture, Snedecor was a talented institution builder. He stimulated demand for statistical consultation through co-hosted seminar series and publications on statistical application in plant-breeding research. The close working relationship he established with agriculture researchers and authorities led to the creation of the Mathematics Statistical Service (later the Statistical Laboratory), which in turn brought resources that enabled ISU to invite world-class visiting scholars, including Fisher and Egon Pearson, for extended visits.

Snedecor was an advocate of Fisher's inferential approach, but he repackaged it in a form that was easier to teach, apply, and institutionalize. His book *Statistical Methods* (1937), written in a more detailed and accessible style, is widely credited with popularizing Fisher's work in the United States. In fact, the book was so accessible that an obituary of Snedecor noted that "the great majority of workers in noisy sciences loved the Snedecor 'cookbook.'" (Kempthorne, 1974, p.320).

On the other side, Neyman parted ways with Egon Pearson, who "lost interest" in the way Neyman was taking (Pearson, as cited in Lehmann, 2011, p. 42). Before the war broke out, Neyman accepted an appointment to the mathematics department at the University of California, Berkeley, in 1938. As his student Lehmann recalled, the department chair who made the appointment likely got more than he had hoped for, as Neyman's ambition was to establish an independent statistics department (Lehmann, 2008). Over the next sixteen years, Neyman worked tirelessly to pursue this goal and finally succeeded in 1955. Lehmann describes Neyman's struggle as follows:

> The struggle to convert a one-man appointment as professor of mathematics into a substantial separate department of statistics did not, of course, take place in a vacuum. It required the acquisition of a faculty, the associated office and laboratory space, a corresponding expansion of the course pro- gram, and, as justification for such an enterprise, an increased enrollment of students taking these courses. Neyman not only carried out these tasks with great skill and unflagging energy, but he also expanded the research program and the resulting financial support for the work of the group. In addition, through symposia and a new series of publications, he created a national and international reputation for his laboratory. (Lehmann, 2008, p. 95).

In addition to his skill and energy for institution building, Neyman's theory-oriented research program readily generated PhD thesis topics, as many formal extensions and refinements could be pursued within the framework. This made it particularly well suited to graduate training and disciplinary reproduction, creating a steady PhD pipeline that later filled "the upper echelons of North American Statistics and beyond, for a decade or more" (Stephen Stigler, cited in Behseta & Kass, 2024, p. 195).

In the United States, clusters of statisticians began to emerge before the war, but the majority of statistical departments were established only after the war (Agresti & Meng, 2012). Prior to the war, only three American universities had formally established academic departments of statistics—Johns Hopkins University, University of Pennsylvania, and George Washington University. In the United Kingdom, the Department of Applied Statistics at University College London (founded by Karl Pearson in 1911) was the only statistics department before the war. As an emerging discipline, statistics departments typically began with graduate programs only, oriented toward specialist training rather than broad instructional demand. In brief, the pre-war capacity for statistical training was quite limited, and the institutional pipeline required for mass undergraduate teaching did not emerge until after the war.

## How WWII and the Cold War Triggered a Statistical Training Boom—and a Supply Crisis

World War II brought statistics to the limelight. During the war, statisticians worked alongside scientists and engineers on a wide range of tasks, from improving weapon production to planning strategic bombing. The war effort demonstrated the power of statistical analysis and had a profound impact on the development of statistics as a discipline. Renowned statistician David Cox recalled:

> World War 2 had the effect of a massive increase in interest in academic statistics, in Cambridge, Oxford, Imperial in the UK and Harvard, Berkeley, Columbia, Stanford, etc. In my own case it is highly unlikely that I would have become a statistician without the War, and I am one of many. (Cited in Agresti, 2021, p. 690)

The success and visibility of statisticians' participation in the war effort spurred post-war demand for trained statisticians, as both public and private sectors sought to improve operation through statistical analysis (National Research Council [NRC], 1947; UN Statistical Commission, 1949; E. S. Pearson, 1959).

Academia was another major driver of demand for trained statisticians, as government policy channeled large amounts of resources into science education and scientific research. With $5.5 billion in funding for higher education tuition, the GI Bill (the Servicemen's Readjustment Act of 1944) brought more than two million veterans to college and university campuses (Olson, 1973). According to Pulitzer prize winner Edward Humes (2006), the GI Bill supported the education of approximately 450,000 engineers and 91,000 scientists, and the resulting influx of students placed severe strain on existing training capacity.

Government funded research further intensified this pressure. Federal support through the National Science Foundation and agencies such as DARPA (Defense Advanced Research Projects Agency) expanded rapidly during the Cold War, driving demand for statistical analysis across a wide range of scientific and technical fields. The National Defense Education Act (NDEA) of 1958, passed soon after the Soviet launch of Sputnik, explicitly linking national security to science and technology education. NDEA prioritized mathematics, science, engineering and modern foreign languages, providing loans to colleges students and fellowships for graduate training in these areas.[3] Under NDEA, federal funding for higher education research and development grew from $138 million in 1953 to over $1.6 billion by 1970.[4]

Natural sciences and engineering were not the only disciplines affected by these policies. Fields closely aligned with Cold War priorities of manpower planning—such as education, psychology, sociology—also experienced rapid growth in doctorate production and federal funding between 1950 and 1970.[5,6] Figures 4-6 illustrate the magnitude of this expansion in total U.S. postsecondary enrollment, research doctorates awarded, and journal articles indexed in OpenAlex.

The surge in demand resulted in a serious shortage of statisticians with adequate training. The magnitude and duration of the shortage are evident in the titles of contemporary publications on both sides of the Atlantic. In 1947, the NRC of the United States published a report titled "Personnel and Training Problems Created by the Recent Growth of Applied Statistics in the United States" (NRC, 1947). As late as 1959, the American Statistical Association was still holding symposia to discuss "the chronic and growing problem of the shortage of statisticians" (Riley & Futransky, 1960, p. 25). In the United Kingdom, the Royal Statistical Society issued the

---

[3] https://www.congress.gov/bill/85th-congress/house-bill/13247/text

[4] https://ncses.nsf.gov/pubs/nsf26304/table/1

[5] National Center for Science and Engineering Statistics (NCSES). Survey of Earned Doctorates: Doctorate Recipients from U.S. Universities. National Science Foundation.

[6] National Center for Science and Engineering Statistics (NCSES). Survey of Federal Funds for Research and Development (Annual R&D Obligations). National Science Foundation.

“Report on the Teaching of Statistics in Universities and University Colleges” in 1947 and formed a Committee on the Supply and Demand for Statisticians in 1956 to address “the general shortage in the supply of trained statisticians” (E. S. Pearson, 1959, p. 47). The concern in these reports extended beyond elite researchers to the availability of instructors and applied practitioners needed to meet rapidly expanding instructional and research demands.

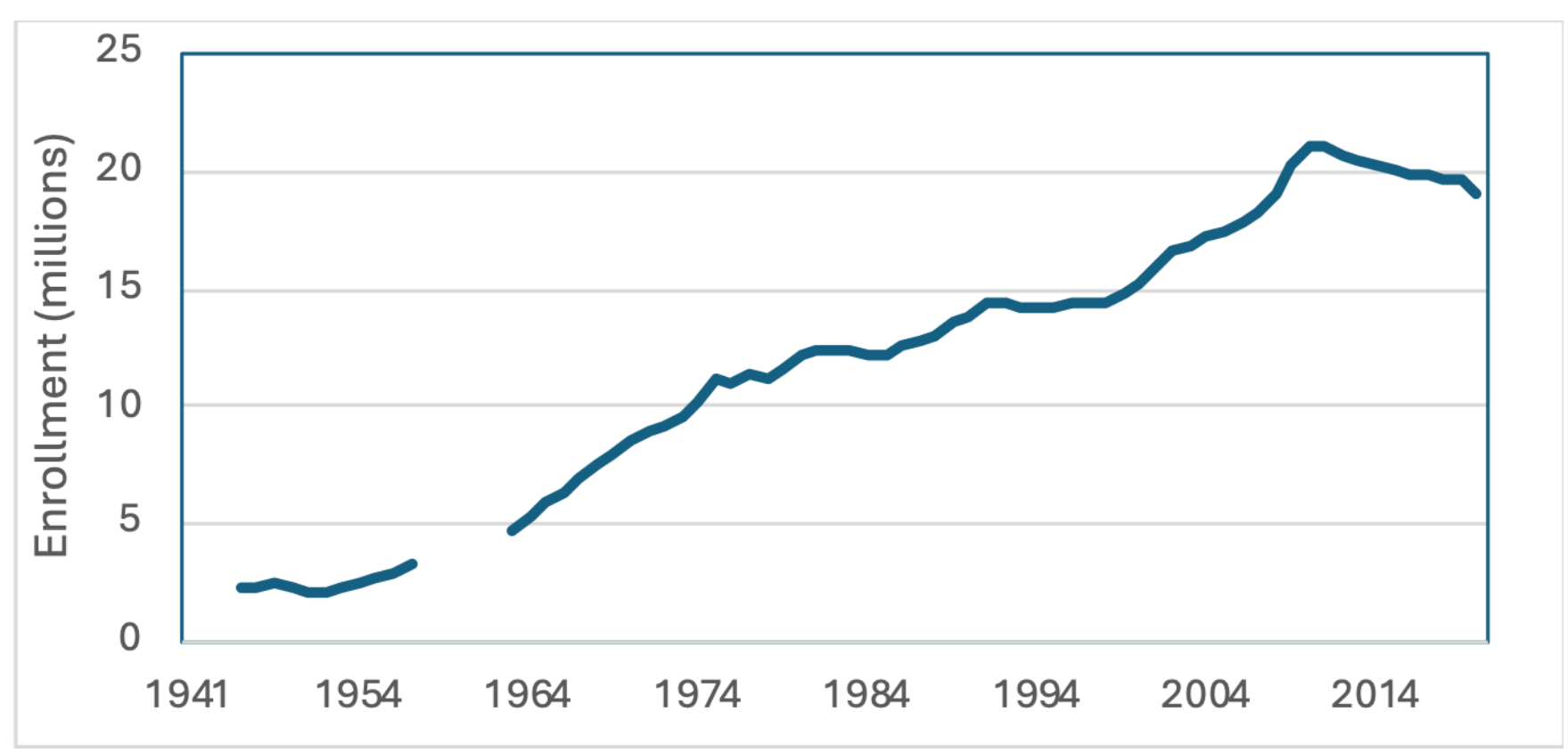

Figure 4 US Postsecondary enrollment

Source: NCES Digest.

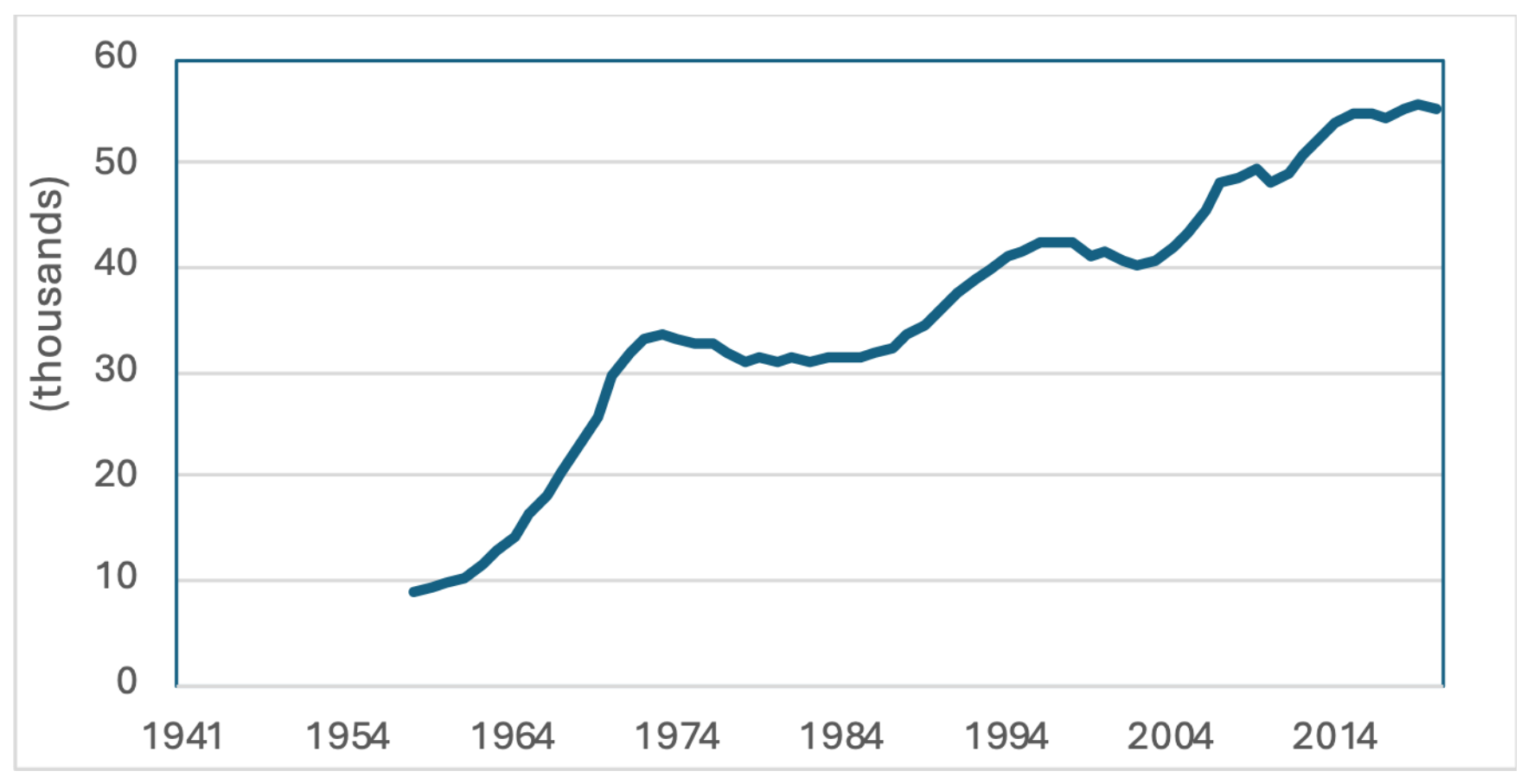

Figure 5 US Research doctorates awarded

Source: NSF NCSES Survey of Earned Doctorates

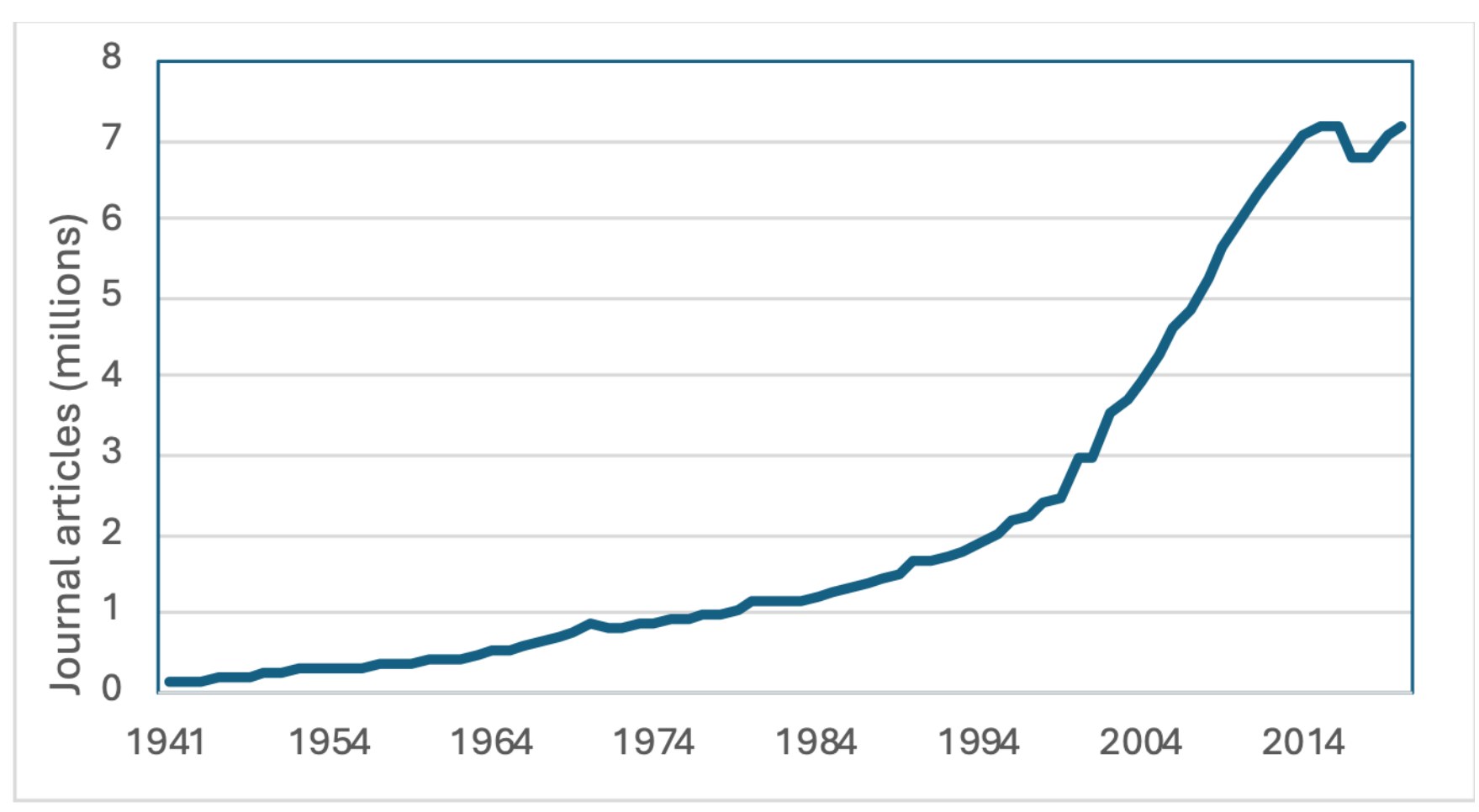

Figure 6 Journal articles indexed in OpenAlex (type = article)

On the supply side, there were few established departments of statistics before the war, and curriculum and professional qualifications were still being developed when the war broke out (Kendall, 1948; Hotelling, 1948; NRC, 1947; E. S. Pearson, 1959; RSS, 1947). Prominent statisticians formed research clusters at major universities after the war, but it often took more than a decade for these clusters to mature into independent departments, requiring the development of faculty, curricula, and training pipelines (Agresti & Meng, 2012). Because of the long and resource-intensive process of building the training infrastructure, the supply of trained statisticians lagged far behind rapidly expanding demand.

While statisticians struggled to scale up training to meet rapidly expanding demand from academia and other sectors, applied programs increasingly relied on self-taught instructors to teach statistics. Harold Hotelling, who played an important role in the establishment of several major U.S. statistics departments in the post-war era, complained forcefully about the state of statistical instruction in the United States:[7]

> The major evil is that those attempting to teach statistical method are all too often not specialists in the subject… There results a widespread teaching of wrong theories and inefficient methods. Students…equipped with the skill that results from careful drilling in methods that ought never to be used. Some of these same students are encouraged and assisted to become college and university teachers of statistics without ever making thorough-going studies of the fundamentals of the subject… Through the method of selection of teachers in general use, and through textbooks written by individuals of this type, there is a perpetuation of obsolete ideas and unsound methods…All this does not mean that any considerable number of people teaching statistics are unworthy or objectionable members of the academic community…The only trouble is that they are teaching a subject in which they are not specialists, and which progresses so fast that only specialists can keep up with it… The uncritical character of the teaching is reflected in the long line of textbooks written by teachers who have not made

[7] In another equally critical address, Hotelling described in general but vivid terms how instruction in non-statistical programs was initiated and sustained (Hotelling, 1940).

> any genuinely fundamental study of statistics, but pass on to students in a magisterial fashion what was passed on to them… It is no wonder that these textbooks, copied from each other, contain increasing accumulations or errors… (Hotelling, 1948, pp. 101-105).

Elsewhere, the situation was described in similarly stark terms: the "proliferation of statistics courses and programs after World War II and a dearth in the number of trained and qualified users of statistics" had reached a "crisis" level (Zieffler, et al., 2017. p. 40). Commenting in his Presidential Address at the 1950 Annual Meeting of the American Statistical Association, Sam Wilks acknowledged the contribution of self-taught instructors in "helping to satisfy immediate instructional needs", while also emphasizing the resulting quality problems in statistical education:

> As a whole and as it now stands, our statistical education structure is something like a group of temporary barracks which have been built in a hurry, built in piece-meal fashion with very little design and built on poor foundations. (Wilks, 1951. p.3)

Under these conditions, reliance on routinized procedures and standardized rules increasingly substituted for deep statistical expertise and contextual judgment.

Of course, as many statistics departments were established in the two decades after World War II, it would be naïve to ignore the issue of resource allocation in such debates. Growing clusters of statisticians on campus inevitably sought to establish independent statistics departments, but this often raised tensions with other departments that already offered statistics courses as part of their curricula. These courses were not only pedagogical sites but also sources of enrollments, teaching positions, and disciplinary jurisdiction.

For example, recounting Neyman's struggle to establish the Statistics Department at UC Berkeley, Lehmann noted that:

> …elementary statistics courses were being taught in economics, education, forestry, psychology, sociology, and a number of other disciplines. And the instructor was often the most junior member of the department, who might have little back-ground in statistics. Despite these shortcomings, the other departments were very reluctant to give up their courses. For one thing, they argued, they could motivate their students better by basing the instruction on examples from their particular subject matter. (Lehmann, 2008, p. 96.)

A remarkable institution builder, Neyman managed to centralize the teaching of lower-division statistics courses across UC Berkeley by employing qualified graduate students from other departments as teaching assistants for these courses (Lehmann, 2008, p. 96). Such tensions between fledgling statistics faculty and other departments offering statistical courses were common in the histories of statistical departments (see the volume edited by Agresti and Meng, 2012).

## The Emergence of the NHST Black Box

The story of NHST is a vivid illustration of the quandary of fact builders that Latour described in *Science in Action* (1987). Both Fisher and Neyman-Pearson sought to promulgate their own theoretical frameworks, but as their ideas began to circulate beyond the small communities in which they were developed, both gradually lost control over how those ideas were understood and used. As selected elements of their work spread through expanding scientific networks, they were translated, simplified, and remixed by users operating under very different institutional and pedagogical constraints.

As a discipline, statistics in the postwar era was characterized by rapid growth as well as contention and division. Alongside the rivalry between Fisher's empirically driven approach and Neyman's orientation towards mathematical abstraction, statisticians also found themselves in ongoing disputes with domain experts in other departments over who was best equipped to teach statistics. These internal and external tensions unfolded simultaneously, shaping the conditions under which particular statistical practices were selected, simplified, and stabilized.

Out of the chaos emerged NHST — the offspring of the "forced marriage" between the Fisherian approach and Neyman-Pearson's decision-theoretic method (Gigerenzer, et al., 1989, p. 106). It is almost impossible to pinpoint when and where NHST first appeared, in part because the two schools used the same term *p*-value in subtly different ways that escaped the attention of most readers. As reformist statistician Greenland acknowledges,

> …the statistics literature up to the highest levels displays descriptions and definitions founded on jargon that is inconsistent across sources and which violate ordinary language meanings. These problems require far more expertise to sort out than ordinary users could be reasonably expected to have. (Greenland, 2019, p. 106)

As the technically demanding original texts passed through successive rounds of translation by textbook authors and instructors, confusion and misconception were not only likely but structurally reinforced. Huberty's review (1993) of popular statistics textbooks suggests that a confusing mixture of terms had already appeared in textbooks before the war. Textbook authors often presented hypothesis testing in terms of one framework while selectively incorporating elements from the other, without acknowledging their distinct origins or logical incompatibilities. Gigerenzer (2004) similarly associated the development and spread of NHST with textbooks written by domain experts in applied sciences. Although these textbook authors rarely articulated the steps as explicitly as present-day NHST tutorials, they freely mixed concepts and inserted their own interpretations of the *p*-value—many of which persist as the most common misconceptions about hypothesis testing today.

NHST's appeal to users of statistical tools is easy to see. It selectively draws from Fisher's and Neyman-Pearson's frameworks what applied researchers and decisionmakers under pressure find useful, while discarding what is too technically demanding or too subtle to sustain at scale. Fisher's test was effectively dismembered at the cutoff value: the part where a *p*-value less than or equal to 0.05 is deemed "significant" was retained, but the "inconclusive when $p>0.05$" part was dropped—likely because it does not produce strong or decisive conclusions.

Neyman and Pearson's method received similar treatment. The clearly specified alternative hypothesis—the engine of Neyman-Pearson's original innovation—is rarely feasible on genuine research frontiers, and was therefore replaced by a tautological alternative often stated simply as "there is an effect", without further specification. With this decapitation, the NP framework collapses back into a single-hypothesis test. Moreover, because researchers seldom have sufficient information to determine and compare the costs of Type I and Type II errors, the task of optimizing error rates was omitted altogether. In its place, the conventional 0.05 became standard, with 0.01 or 0.001 occasionally invoked to signal greater "confidence".

The result was the loss of Neyman-Pearson's original rigor, leaving behind only an automatic decision rule that classifies results as significant or not based on an arbitrary cutoff. Under conditions of institutional pressure, this hybrid can appear to offer the best of both worlds: black-and-white answers, apparent rigor, and decisive conclusions—without having to wait for the long run.

As demand for statistical analysis grew rapidly while training lagged, a decontextualized cookbook approach to instruction emerged as the standard pedagogy in applied statistics. Integral to NHST, this pedagogical approach strips away the context of data generation and modeling assumptions—the very elements that link sample estimates to the population parameter under study (see the "Sampling Distribution" Section). By omitting the step of examining these assumptions, users of statistical tools are effectively asked to take a leap of faith and assume that all modeling assumptions are met, even when they may be violated to an important extent (Greenland, 2025).

Sidestepping modeling assumptions, NHST renders interpretation largely irrelevant and turns hypothesis testing into a mechanical procedure:

1. State the null hypothesis as there is no effect, difference, or association
2. Reject the null if $p \leq 0.05$; otherwise accept the null

This mechanical routine concludes with labeling results as either significant or insignificant. Unaware of the peculiar origin of the term "significant" (Shafer, 2020), many NHST users treat the label as self-explanatory and perceive little need for interpretation. The dichotomous classification becomes the focal point, while questions about the magnitude of the effect and its dependence on context fade into the background. If the *p*-value falls below the cutoff, the unspecified alternative hypothesis is accepted and a discovery is declared, without warning about the arbitrariness of the threshold or the possibility that the effect may be substantively trivial. If the *p*-value exceeds the cutoff, researchers often conclude that there is "no effect", unaware that the logical basis for accepting the null is absent. It can be said that NHST works partially because it hides fragility. In practice, this concealment is reinforced by a tendency to force a deterministic frame on phenomena whose nature requires probabilistic interpretation (Ting & Greenland, 2024). Under these conditions, hypothesis testing was increasingly reduced to a routine whose internal assumptions and interpretive demands were sealed off from view, leaving behind a standardized output that could circulate independently of the expertise required to produce it.

As all sectors grew rapidly after the war, meeting the demand for manpower became the most pressing problem. Under these conditions, there was little time for the average user of statistical tools to build up the solid understanding required by either the Fisherian or Neyman-Pearson framework. With its promise of ease and apparent power, NHST filled this gap and soon came to dominate many branches of the applied sciences (Fidler, 2005). Against this backdrop, generations of students were taught to equate statistical analysis with feeding data into statistical software programs and applying standard labels to the output. There was no need to understand what the procedures did or how they worked: NHST turns statistics into a black box, enabling the mass adoption and mass production of quantitative research.

## NHST as a Technology of Institutional Massification: Procedural Self-Sufficiency

One major development in postwar America was institutional massification based on the vision laid out in Vannevar Bush's "Science—the endless frontier", a report submitted to President Truman in 1945. Seizing the moment when American trust in science was at its historical peak, this document—often hailed as the "Magna Carta of America Science" (Thorp, 2020, p. 227)—argued powerfully that basic scientific research was a national strategic resource essential to public health, national security, and economic welfare. Bush emphasized the importance of maintaining the momentum of wartime research and called for permanent government support for the large-scale development of scientific knowledge and manpower:

> The Government should accept new responsibilities for promoting the flow of new scientific knowledge and the development of scientific talent in our youth. These responsibilities are the proper concern of the Government, for they vitally affect our health, our jobs, and our national security. It is in keeping also with basic United States policy that the Government should foster the opening of new frontier and this is the modern way to do it. (Bush, 1945, pp. 4-5)

Bush's skillful persuasion paved the way for the establishment of National Science Foundation and the sustained injection of federal funds into higher education and scientific research. Under this national zeitgeist, higher education expanded rapidly, and applied science departments, programs, and research centers proliferated across university campuses (Loss, 2017).

However, the vision of mass participation in science and technology cannot be realized without a scalable infrastructure—one that can rapidly expand the supply of trained personnel and efficiently coordinate activity as the system grows exponentially in size. In the case of statistics, the prewar elite system of training was not to the task, because its pedagogical and credentialing arrangements did not scale. Elite training emphasized deep understanding of data-generation processes and domain-specific causal reasoning, which required close apprenticeship under an expert statistician. This form of training was slow and could not be expanded on demand. A second bottleneck lay in the epistemic labor itself: the work of deliberating thick, context-dependent expert judgments, negotiating interpretive disagreements, and accounting for uncertainty. Under conditions of rapid expansion, both the time-intensive apprenticeship model and the costs of sustained expert deliberation severely limited the scalability of traditional statistical pedagogy and credentialing.

From this vantage point, the often-cited complaints about NHST's cookbook pedagogy and technical slippages take on new meaning. Rather than unfortunate side effects, they are features of a social technology demanded by the pressure to rapidly scale up a country's scientific participation and production. Rapid expansion replaces face-to-face negotiation with indirect cooperation facilitated by standard protocols and procedures, making defensibility and auditability central conditions for governing at scale. In this sense, inferential procedures that could be specified in advance and executed as routine sequences of formal steps—what might be described as programmable inference[8]—were particularly well suited to the demands of rapidly expanding scientific institutions. As growing number of students and papers overwhelmed the supply of true experts and pressure for making quick decisions mounted, NHST emerged as a distributed and largely uncoordinated response to these pressures and uneven expertise.

By purging context and replacing judgment with standardization, NHST reduces the demand for specialized statistical expertise, thereby enabling universities to rely on interchangeable, non-expert instructors for large scale teaching. For instructors overwhelmed with surging teaching loads, the routinized procedures convey a sense of rigor while remaining easy to teach and assess. In research, NHST functions as an effective credentialling and signaling device that differentiates insiders from outsiders, while its dichotomization of results makes decisions defensible and auditing straightforward.

Because NHST replaces judgment with procedure, there is no need for repeated renegotiation of epistemic assumptions. Coordination costs are therefore minimal and largely independent of the number of participants. NHST also operates as a highly efficient adjudicating and sorting device for allocating resources across the scientific system. A grant officer at a funding agency may prioritize long-term impact, a tenure-track researcher may focus on projects that yield results quickly, and a journalist may attend to studies with immediate relevance or reader appeal. In such situations—where evaluative criteria differ and judgment is costly—it is sensible to delegate decisions to a procedure already trusted by those being evaluated.

[8] I thank Sander Greenland for suggesting the characterization of formal statistical procedures as "programmable" in correspondence, which helped clarify the connection between proceduralization and institutional scalability discussed here.

Across teaching, research, and administration, NHST scales well because it is *procedurally self-sufficient*: the procedure itself performs the work of adjudication and persuasion, largely independent of local context. In this sense, a procedural epistemology takes shape, in which trust attaches to claims whose production processes conform to standardized procedures.

To appreciate the power NHST's authority and procedural self-sufficiency, consider the following phenomenon. Recall that NHST users typically deploy the *p*-value as a decision device in a way that is logically equivalent to flipping a coin to decide whether to bring an umbrella in changeable weather. While nobody would expect a coin flip to also indicate the probability that the resulting decision is correct, this is precisely what happens once one enters the research context: NHST papers routinely claim discoveries of substantial impact with a high degree of certainty because *p*-values fall below a conventional threshold. In this setting, the *p*-value is made to do two things at once—both to decide outcomes and to stand in for the probability that the decision is right. As Gelman (2016) notes, this move departs sharply from ordinary standards of reasoning. What requires explanation is not a simple misunderstanding, but a striking shift in epistemic expectations across contexts. The most plausible account is that the NHST procedure carries such institutional authority that it overrides otherwise stable intuitions about what decision rules can and cannot tell us.

For better or worse, thanks to the procedural self-sufficiency it affords, NHST rose to the challenge of postwar institutional massification in the United States, turning universities into apparatuses for the mass production of science manpower. Institutional massification legitimized NHST, and NHST in turn enabled institutional massification. As Latour put it, "understanding what facts and machines are is the same task as understanding who the people are" (1987, p. 140).

## Discussion

Drawing on statistical literature and research on the history of statistics, this paper has explained how NHST—and the concept of statistical significance—came to achieve its present dominance. By tracing the genealogy of NHST and situating its development within the historical condition of postwar expansion, the analysis shows that NHST emerged from theoretical contention and institutional disorder in statistical training, and that it stabilized because it offered a social solution to the most pressing problem of the period: institutional massification.

This sociological approach helps resolve a longstanding puzzle: why a method widely decried by statistical luminaries nevertheless travels and endures so well, despite decades of educational and reform efforts? The answer is that what NHST solved was not the technical problem of statistical inference, but the social problem of rapidly bringing a massive number of people into scientific practice. As the analysis has shown, the massification of science relies on epistemic thinning that gradually transformed the original ideas of Fisher and Neyman-Pearson into a mechanical routine that is procedurally self-sufficient. In this sense, NHST became the obligatory passage point in many scientific fields precisely because it is largely indifferent to users' epistemic commitments: the procedure itself performs the work of adjudication and persuasion. This procedural self-sufficiency makes NHST highly redeployable, which helps explain why it travels so effectively and continues to attract new users wherever institutional massification takes place.

The broader message is that, for a technically demanding practice such as statistical inference, massification tends to require participation technologies that drift away from their epistemic origins and give rise to technically questionable practices. As long as such drift does not produce immediately observable or verifiable negative effects—and as long as its costs are not internalized, unlike in domains such as driving—procedural self-sufficiency is likely to dominate other considerations.

In this sense, history may already be repeating itself. For decades, Bayesian statistics has been proposed as an alternative that avoids many of NHST's well-known limitations, and its use has been on the rise as computing power continues to grow. Yet there are signs that the Bayes Factor is increasingly taking on a role analogous to that of the *p*-value, serving as a device for the automation of decisions rather than as a tool for substantive inference (Muradchanian, et al., 2024). This trajectory does not invite much optimism if reform efforts focus on technical substitution alone.

If improved teaching is unlikely to suffice under conditions of mass participation in science, is there any hope for statistical reform? The answer is not entirely pessimistic. Once the problem is recognized as a social one, reform efforts can be understood to require a social strategy alongside a technical strategy (Sismondo, 2010). More specifically, because NHST's dominance stems in part from users mistaking its social function for a technical one, history and sociology can be mobilized to make students and practitioners aware of what NHST does institutionally, and what they gain—and lose—when they rely on it. This paper is one such attempt.

**Acknowledgments**

I thank Sander Greenland for helpful comments on an earlier version of this manuscript and for pointing out relevant references. Any remaining errors or interpretations are my own.